\documentclass[referee]{raa}            

\usepackage{graphicx,times}             

\begin{document}

   \title{Strongly screening electron capture rates of chromium isotopes in presupernova$^*$
 \footnotetext{$^{*}This~ work~ is ~supported~ in~ part ~by~ the~ National~ Natural~ Science~ Foundation~ of ~China ~under
~grants~ 11565020, \\~10773005, ~and~ the ~Counterpart~ Foundation
~of ~Sanya ~under ~grant ~2016PT43,~ the ~Special~ Foundation~ of~
Science~ and~\\ Technology~ Cooperation ~for ~Advanced~ Academy~
and~ Regional~ of~ Sanya~ under~ grant~ 2016YD28,~ the~
Scientific~Research \\~Staring ~Foundation~ for~ 515~Talented
~Project ~of~ Hainan~ Tropical~ Ocean~ University~ under~ grant~
RHDRC201701,\\~ and ~the ~Natural ~Science~ Foundation~ of ~Hainan~
province~ under~ grant~ 114012.$ } }

   \volnopage{Vol.0 (201x) No.0, 000--000}      
   \setcounter{page}{1}          

   \author{Jing-Jing Liu\inst{1}, Qiu-He,Peng\inst{2}, Liang-Huan Hao\inst{1}, Xiao-Ping Kang\inst{1}, and Dong-Mei Liu\inst{1}}

   \institute{$^{1}$College of Marine Science and Technology, Hainan Tropical Ocean University, Sanya, 572022, China; {\it liujingjing68@126.com}\\
$^{2}$Department of Astronomy, Nanjing University, Nanjing, Jiangshu
210000, China.
   }

   \date{Received~~2014 month day; accepted~~2014~~month day}

\abstract{Taking into account the effect of electron screening on
the electron energy and electron capture threshold energy, by using
the method of Shell-Model Monte Carlo and Random Phase Approximation
theory, we investigate the strong electron screening capture rates
of chromium isotopes according to the linear response theory
screening model. The strong screening rates can decrease by about
40.43\% (e.g., for $^{60}$Cr at $T_9=3.44,Y_e=0.43$). Our
conclusions may be helpful to the researches of supernova explosion
and numerical simulation. }

   \authorrunning{Jing-Jing. Liu }            
   \titlerunning{Strongly screening electron capture rates of chromium isotopes in presupernova}  

   \maketitle

%
%
\section{Introduction}

At the presupernova stage, beta decay and electron capture on some
neutron-rich nuclei may play an important roles in determining the
hydrostatic core structure of massive presupernova stars, thereby
affect the subsequent evolution during the gravitational collapse
and supernova explosion phases (e.g., Dean et al. 1998; Caurier et
al. 1999; Juodagalvis et al. 2010; Liu 2013a; 2014; 2016a; 2016b;
2017). For example, the beta decay (electron capture) strongly
influences the time rate of change of the lepton fraction (e.g., the
time rate of change of electron fraction $\dot{Y_e}$) by increasing
(decreasing) the number of electrons. Some isotopes of iron,
chromium, and copper can also make a substantial contribution to the
overall changes in the lepton fraction (e.g., $\dot{Y_e}$), electron
degenerate pressure, and entropy of the stellar core during its very
late stage of evolution. Many of these nuclei could be appropriately
tracked in the reaction network in the stellar evolution
calculations. The lepton fraction (e.g., $\dot{Y_e}$) is bound to
lead to an unstoppable process of gravitational collapse and
supernova explosion.

Some research shows that the EC of iron group nuclei (e.g., iron and
chromium isotopes) are very important and dominate for supernova
explosions (e.g., Aufderheide et al. 1990, 1994; Dean et al. 1998;
Heger et al. 2001; ). In the process of presupernova evolution,
chromium isotopes are a very important and crucial radionuclide.
Aufderheide et al. (1994) detailed investigated the EC and beta
decay for these nuclei in presupernova evolution. They found that
the EC rates of these chromium isotopes can be of significant
astrophysical importance by controlling the electronic abundance.
Heger et al. (2001) also discussed weak-interaction rates for some
iron group nuclei by employing shell model calculations in
presupernova evolution. They found that electron capture rates on
iron group nuclei would be crucial for decreasing the electronic
abundance ($Y_e$) in stellar matter.

On the other hand, in the process of presupernova evolution of
massive stars, the Gamow-Teller transitions of isotopes of chromium
play a consequential role. Some studies shown that $\beta$-decay and
electron capture rates on chromium isotopes significantly affect the
time rate of change of lepton fraction ($\dot{Y_e}$). For example,
Nabi et al. (~\cite{Nabi16}) detailed the Gamow-Teller strength
distributions, $\dot{Y_e}$, and neutrino energy loss rates for
chromium isotopes due to weak interactions in stellar matter.

However, their works did not discuss the problem that electron
screening (SES) would strongly effect on EC. What role does the EC
play in stellar evolution? How does SES influence on EC reaction at
high density and temperature? In order to calculate accurately the
EC rates and screening correction for supernova explosion and
numerical simulation, in this paper we will detailed discuss this
problem.

Based on the linear response theory model (LRTM) and Random Phase
Approximation (RPA), we study the strong screening EC rates of
chromium isotopes in astrophysical environments by using the
Shell-Model Monte Carlo (SMMC) method. In the next Section, we
discuss the methods used for EC in stellar interiors in the case
with and without SES. Section 3 will present some numerical results
and discussions. Conclusions follow in Section 4.

\section{The EC rates in the process of stellar core collapse}
\label{sect:The}

\subsection{The EC rates in the case without SES}
For nucleus $(Z, A)$, we calculate the stellar EC rates, which is
given by a sum over the initial parent states $i$ and the final
daughter states $f$ at temperature $T$ and it is written by (e.g.,
Fuller et al. 1980, 1982)
\begin{equation}
  \lambda_{k}=\sum_i \frac{(2J_i+1)e^{\frac{-E_i}{kT}}}{G(Z, A, T)} \sum_{f} \lambda_{if}
\label{eq:001}
\end{equation}
here $J_i$ is the spin and $E_i$ is excitation energies of the
parent states, the nuclear partition function $G(Z, A, T)$ has been
discussed by Aufderheide et al. (1990, 1994). $\lambda_{if}$ is
named as the rates from one of the initial states to all possible
final states.

Based on the theory of RPA, the EC rates is closely related to cross
section $\sigma_{ec}$, and we can written by (e.g., see detailed
discussions in Dean et al. 1998; Caurier et al. 1999; Juodagalvis et
al. 2010)
\begin{equation}
  \lambda_{if}=\frac{1}{\pi^{2}\hbar^{3}}\sum_{if}\int^{\infty}_{\varepsilon_{0}}p^2_e
  \sigma_{ec}(\sigma_e,\sigma_i,\sigma_f)f(\sigma_e,U_F,T)d\varepsilon_e
\label{eq:002}
\end{equation}
where $\varepsilon_0=\max(Q_{if}, 1)$. The incoming electron
momentum is $p_e=\sqrt{\varepsilon_e-1}$, and $\varepsilon_{e}$ is
the electron energy and the electron chemical potential is given by
$U_{F}$, $T$ is the electron temperature. The energies and the
moments are in units of $m_e c^2$ and $m_e c$ ($m_e$ is the electron
mass and $c$ is the light speed), respectively.

The electron chemical potential is obtained by
\begin{equation}
  n_e=\frac{\rho}{\mu_e} =\frac{8\pi}{(2\pi)^3}\int^\infty_0 p^2_e(G_{-e}-G_{+e})dp_e
\label{eq:003}
\end{equation}
here $\mu_e$, $\rho$ are the average molecular weight and the
density in $\rm{g/cm}^3$, respectively. $\lambda_e=\frac{h}{m_{e}c}$
is the Compton wavelength,
$G_{-e}=[1+\exp(\frac{\varepsilon_{e}-U_{F}-1}{kT})]^{-1}$ and
$G_{+e}=[1+\exp(\frac{\varepsilon_{e}+U_{F}+1}{kT})]^{-1}$ are the
electron and positron distribution functions respectively, $k$ is
the Boltzmann constant. The phase space factor is defined as
\begin{equation}
  f(\varepsilon_{e},U_F,T)=[1+\exp(\frac{\varepsilon_{e}-U_F}{kT})]^{-1}
\label{eq:004}
\end{equation}

According to the energy conservation, the electron, proton and
neutron energies are related to the neutrino energy, and $Q$-value
for the capture reaction (Cooperstein et al. 1984)
\begin{equation}
  Q_{if}=\varepsilon_{e}-\varepsilon_{\nu}=\varepsilon_{n}-\varepsilon_{\nu}=\varepsilon^{n}_{f}-\varepsilon^{p}_{i}
\label{eq:005}
\end{equation}
and we have
\begin{equation}
  \varepsilon^{n}_{f}-\varepsilon^{p}_{i}=\varepsilon^{\ast}_{if}+\hat{\mu}+\Delta_{np}
\label{eq:006}
\end{equation}
where $\varepsilon_{\nu}$ is neutrino energy, $\varepsilon_i^p$ is
the energy of an initial proton single particle state,
$\varepsilon_f^n$ is the energy of a neutron single particle state.
$\hat{\mu}=\mu_{n}-\mu_p$ and
$\Delta_{np}=M_{n}c^2-M_{p}c^2=1.293$MeV are the chemical potentials
and mass difference between neutron and proton in the nucleus,
respectively. $Q_{00}=M_{f}c^2-M_{i}c^2=\hat{\mu}+\Delta_{np}$, and
the masses of the parent nucleus and the daughter nucleus are
corresponding to $M_{i}$ and $M_{f}$; $\varepsilon^{\ast}_{if}$ is
the excitation energies for daughter nucleus at zero temperature
state.

The total cross section in the process of EC reaction is given by
(e.g., Dean et al. 1998; Caurier et al. 1999; Juodagalvis et al.
2010)
\begin{eqnarray}
\sigma_{ec}=\sigma_{ec}(\varepsilon_e)&=&
\sum_{if}\frac{(2J_{i}+1)\exp(-\beta
\varepsilon_i)}{Z_A}\sigma_{fi}(\varepsilon_e)
=\sum_{if}\frac{(2J_{i}+1)\exp(-\beta \varepsilon_i)}{Z_A}\sigma_{fi}(\varepsilon_e) \nonumber\\
&=& 6g^{2}_{wk}\int d\xi(\varepsilon_{e}-\xi)^2 \frac{G^2_A}{12\pi}
S_{\rm{GT}^+}(\xi) F(Z,\varepsilon_e) \label{007}
\end{eqnarray}
where $g_{wk}=1.1661\times 10^{-5}\rm{GeV}^{-2}$ is the weak
coupling constant and $G_A=1.25$. $F(Z, \varepsilon_e)$ is the
factor of Coulomb wave correction.

The total amount of Gamow-teller(GT) strength is $S_{\rm{GT}^+}$
which is by summing over a complete set from an initial state to
final states. The response function $R_A(\tau)$ of an operator
$\hat{A}$ at an imaginary-time $\tau$ is calculated by using the
method of SMMC. Thus, $R_A(\tau)$ is given by (e.g., Dean et al.
1998; Juodagalvis et al. 2010)
\begin{equation}
R_A(\tau)=\frac{\sum_{if}(2J_i+1)e^{-\beta \varepsilon_i}e^{-\tau
(\varepsilon_f-\varepsilon_i)}|\langle f|\hat{A}|i\rangle|^2}{\sum_i
(2J_i+1)e^{-\beta \varepsilon_i}} \label{eq:008}
\end{equation}

The strength distribution is is related to $R_A(\tau)$ by a Laplace
Transform $R_A(\tau)=\int_{-\infty}^{\infty}S_A(\varepsilon)e^{-\tau
\varepsilon}d\varepsilon$ and given by (e.g., Dean et al. 1998;
Caurier et al. 1999; Juodagalvis et al. 2010)
\begin{equation}
S_{GT^+}(\varepsilon)=S_{A}(\varepsilon)=\frac{\sum_{if}\delta
(\varepsilon-\varepsilon_f+\varepsilon_i)(2J_i+1)e^{-\beta
\varepsilon_i}|\langle f|\hat{A}|i\rangle|^2}{\sum_i
(2J_i+1)e^{-\beta \varepsilon_i}} \label{eq:009}
\end{equation}
here $\varepsilon$ is the energy transfer within the parent nucleus,
and the $S_{\rm{GT}^+}(\varepsilon)$ is in units of $\rm{MeV}^{-1}$
and $\beta=\frac{1}{T_N}$, and $T_N$ is the nuclear temperature.

For degenerate relativistic electron gas, the EC rates in the case
without SES are given by (e.g., Dean et al. 1998; Caurier et al.
1999; Juodagalvis et al. 2010)

\begin{equation}
\lambda_{ec}^0=\frac{\ln2}{6163}\int^{\infty}_{0}d\xi
S_{\rm{GT}^+}\frac{c^3}{(m_{e}c^2)^5}\int^{\infty}_{p_0}dp_{e}p^2_e(-\xi+\varepsilon_e)^2
F(Z,\varepsilon_e)f(\varepsilon_e,U_F,T) \label{118} \label{eq:010}
\end{equation}
The $p_0$ is defined as
\begin{equation}
p_0=\left\lbrace \begin{array}{ll}~\sqrt{Q^2_{if}-1}~~~~~~( Q_{if}<-1)\\
                                  ~0 ~~~~~~(\rm{otherwise}).
                             \end{array} \right.
\label{011}
\end{equation}

\subsection{The EC rates in the case with SES}

In 2002, based on the linear response theory model (LRTM) for
relativistic degenerate electrons Itoh et al.(2002) discussed the
effect of the screening potential on EC. The electron is strongly
degenerate in our considerable regime of the density-temperature.
The condition is expressed as
\begin{small}
\begin{equation}
 T\ll T_F=5.930\times10^9\{[1+1.018(\frac{Z}{A})^{2/3}(10\rho_7)^{2/3}]^{1/2}-1\},
 \label{eq.19}
\end{equation}
\end{small}
here $T_{\rm{F}}$ and $\rho_7$ are the electron Fermi temperature
and the density (in units of $10^7\rm{g/cm^3}$).

For relativistically degenerate electron liquid, Jancovici et al.
(1962) studied the static longitudinal dielectric function. Taking
into account the effect of strong screening, the electron potential
energy is written by
\begin{equation}
V(r)=-\frac{Ze^2(2k_{\rm{F}})}{2k_{\rm{F}}r}\frac{2}{\pi}\int_0^\infty
\frac{\rm{sin}[(2k_{\rm{F}}r)]q}{q\epsilon(q,0)}dq,
 \label{eq.20}
\end{equation}
where $\epsilon(q,0)$ is Jancovici¡¯s static longitudinal dielectric
function and $k_{\rm{F}}$ is the electron Fermi wave-number.

The screening potential for relativistic degenerate electrons by
linear response theory is written by (Itoh et al. 2002)
\begin{equation}
D=7.525\times10^{-3}Z(\frac{10z\rho_7}{A})^{\frac{1}{3}}J(r_s,R)~~
(\rm{MeV}) \label{eq.21}
\end{equation}
Itoh et al.(2002) detailed discussed the parameters $J(r_s,R)$,
$r_s$ and $R$. The Eq. (14) is fulfilled in the pre-supernova
environment and is satisfied for $10^{-5}\leq r_s \leq 10^{-1},
~~0\leq R\leq 50$.

The screening energy is sufficiently high enough such that we can
not neglect its influence at high density when electrons are
strongly screened. The electron screening will make electron energy
decrease from $\varepsilon$ to $\varepsilon^{'}=\varepsilon-D$ in
the process of EC. Meanwhile the screening relatively increases
threshold energy from $\varepsilon_0$ to
$\varepsilon_s=\varepsilon_0+D$ for electron capture. So the EC
rates in SES is given by (e.g., Juodagalvis et al. 2010; Liu. 2014)
\begin{eqnarray}
\label{eq.22}
\lambda^{s}_{ec}
=\frac{\ln2}{6163}\int^{\infty}_{0}d\xi S_{GT^+}\frac{c^3}{(m_{e}c^2)^5}\nonumber\\[1mm]
\int^{\infty}_{\varepsilon_s}d\varepsilon^{'}\varepsilon{'}(\varepsilon^{'2}-1)^{\frac{1}{2}}(-\xi+\varepsilon^{'})^2
F(Z,\varepsilon^{'})f(\varepsilon_e,U_F,T)
\end{eqnarray}

The nuclear binding energy will increase due to interactions with
the dense electron gas in the plasma. The effective nuclear Q-value
($\rm{Q}_{if}$), will change at high density due to the influence of
the charge dependence of this binding. When we take account into the
effect of SES, the electron capture Q-value will increase by (Fuller
et al(1982))
\begin{equation}
\Delta Q\approx 2.940\times10^{-5}Z^{2/3}(\rho
Y_e)^{1/3}~~~\rm{MeV}. \label{eq.21}
\end{equation}
Therefore, The Q-value of EC increases from $Q_{if}$ to
$Q_{if}'=Q_{if}+\Delta Q$. The $\varepsilon_s$ is defined as
\begin{equation}
\varepsilon_s=\left\lbrace \begin{array}{ll}~Q_{if}'+D~~~~~~(Q_{if}'<-m_ec^2)\\
                                  ~m_ec^2+D ~~~~~~(\rm{otherwise}).
                             \end{array} \right.
\label{eq.16}
\end{equation}

We define the screening enhancement factor C to enable a comparison
of the results as follows

\begin{equation}
C=\frac{\lambda^{s}_{ec}}{\lambda^{0}_{ec}} \label{eq.21}
\end{equation}



\begin{figure}
\centering
    \includegraphics[width=7cm,height=7cm]{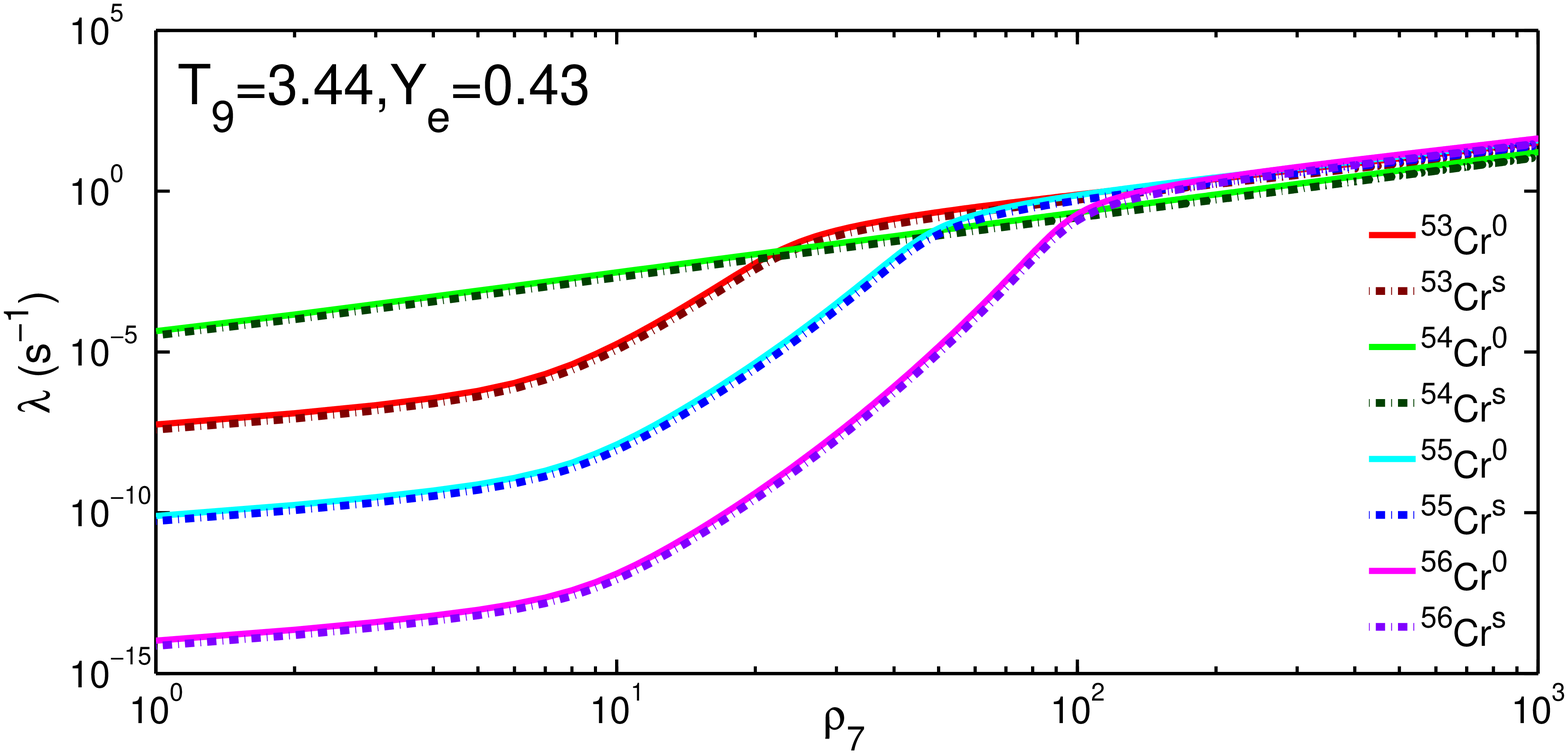}
    \includegraphics[width=7cm,height=7cm]{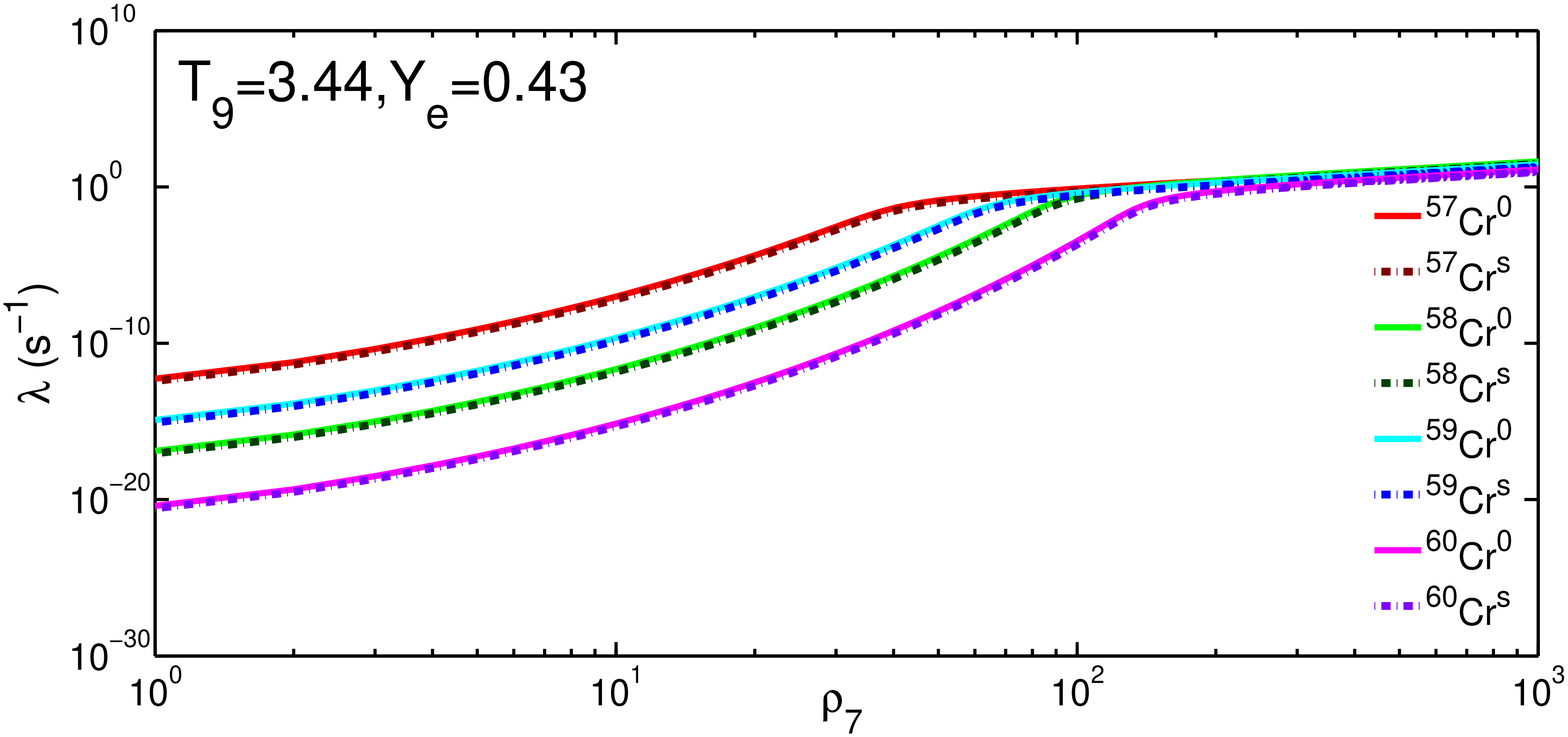}
    \includegraphics[width=7cm,height=7cm]{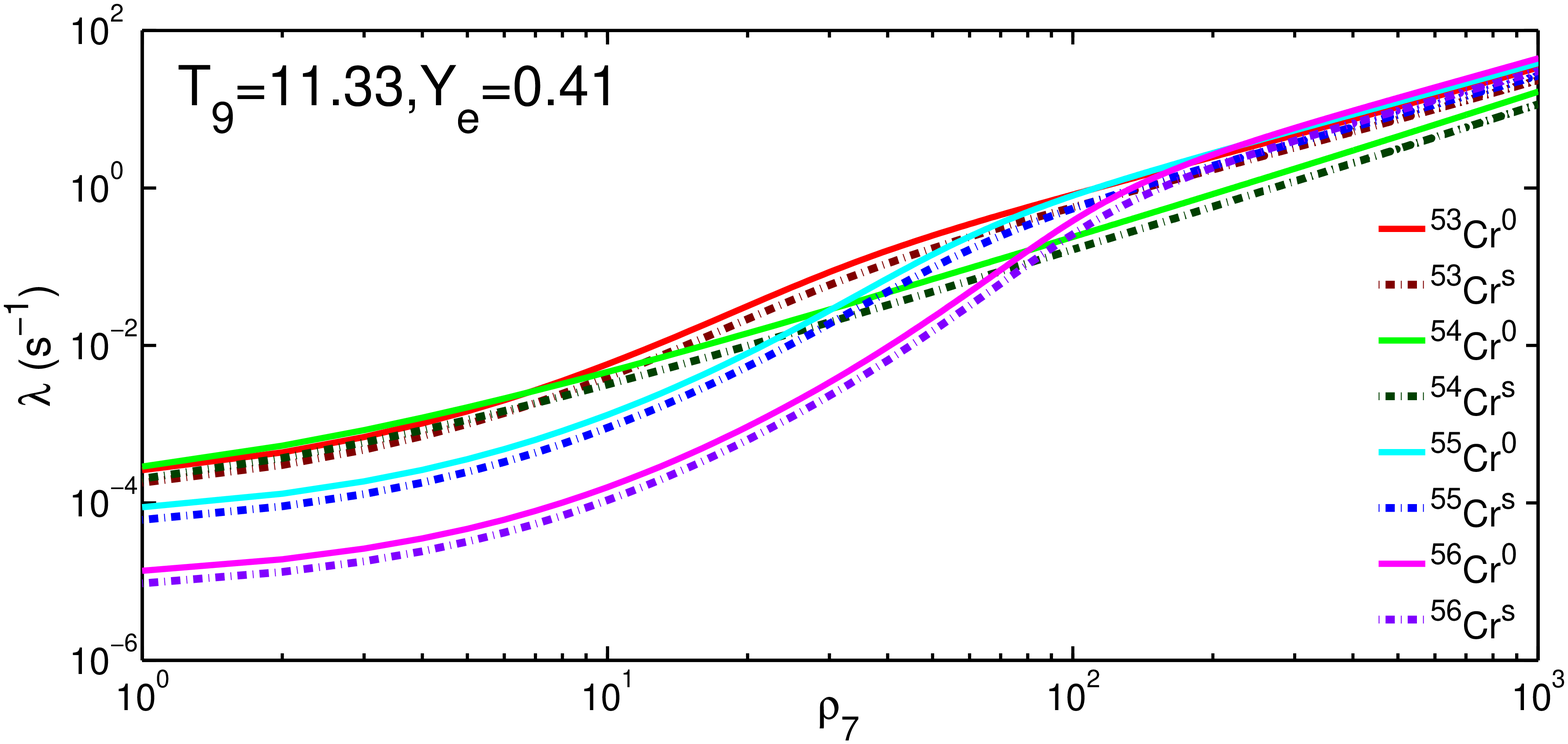}
    \includegraphics[width=7cm,height=7cm]{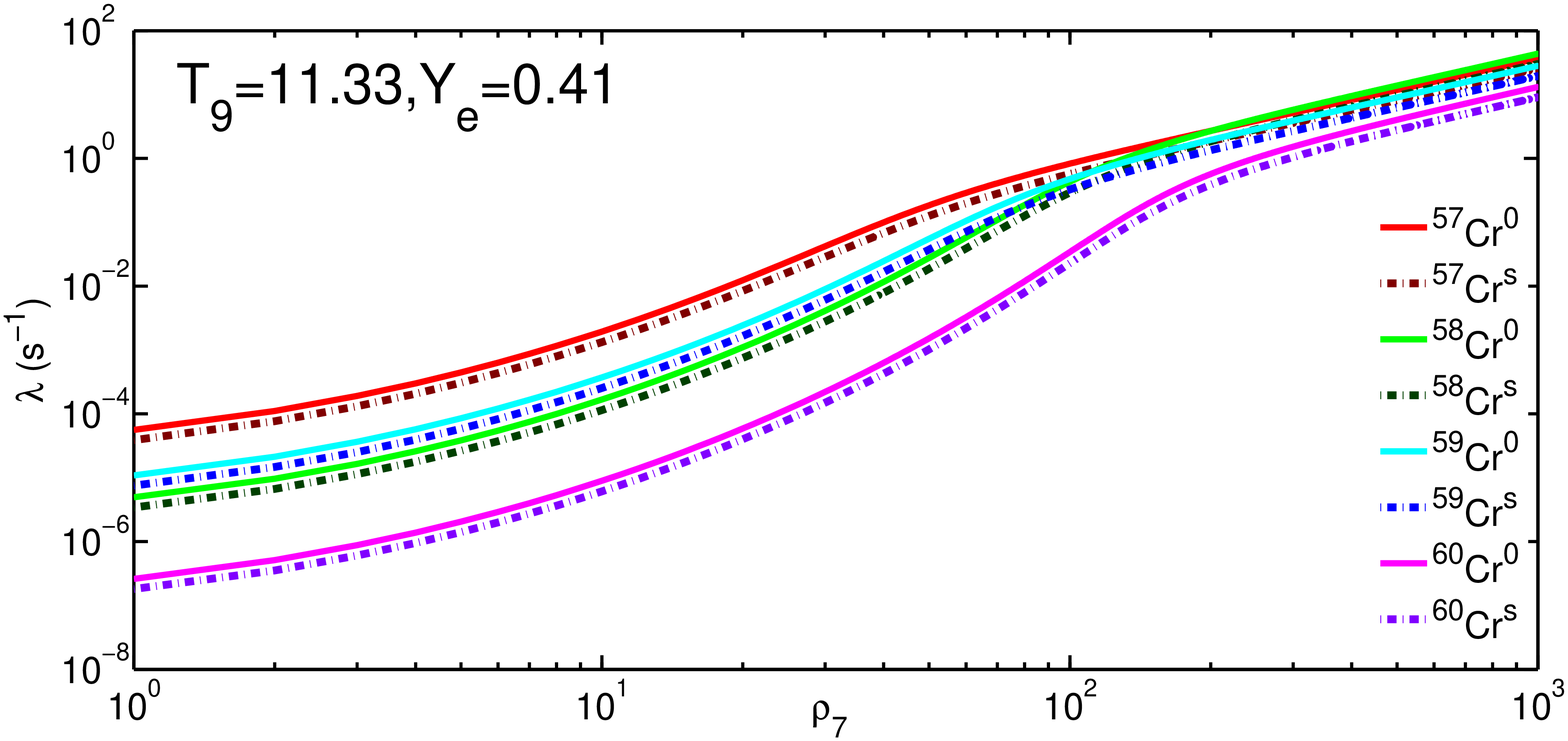}
   \caption{The no SES and SES rates corresponding to solid and dotted line for chromium isotopes as a function of the density $\rho_7$ at the
temperature of $T_9=3.44, Y_e=0.43$ and $T_9=11.33, Y_e=0.41$.}
   \label{Fig:1}
\end{figure}

%

\begin{figure}
\centering
   \includegraphics[width=7cm,height=7cm]{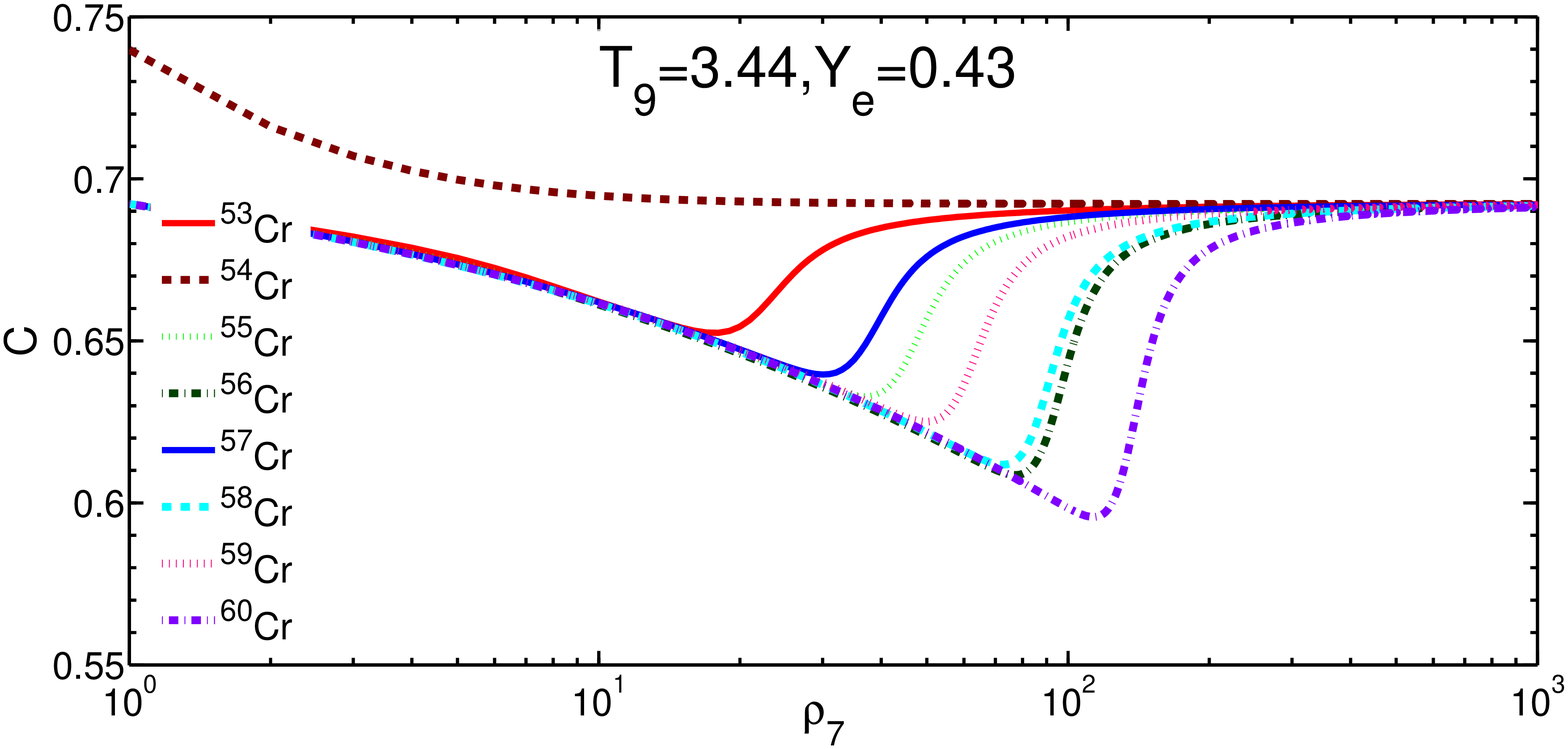}
    \includegraphics[width=7cm,height=7cm]{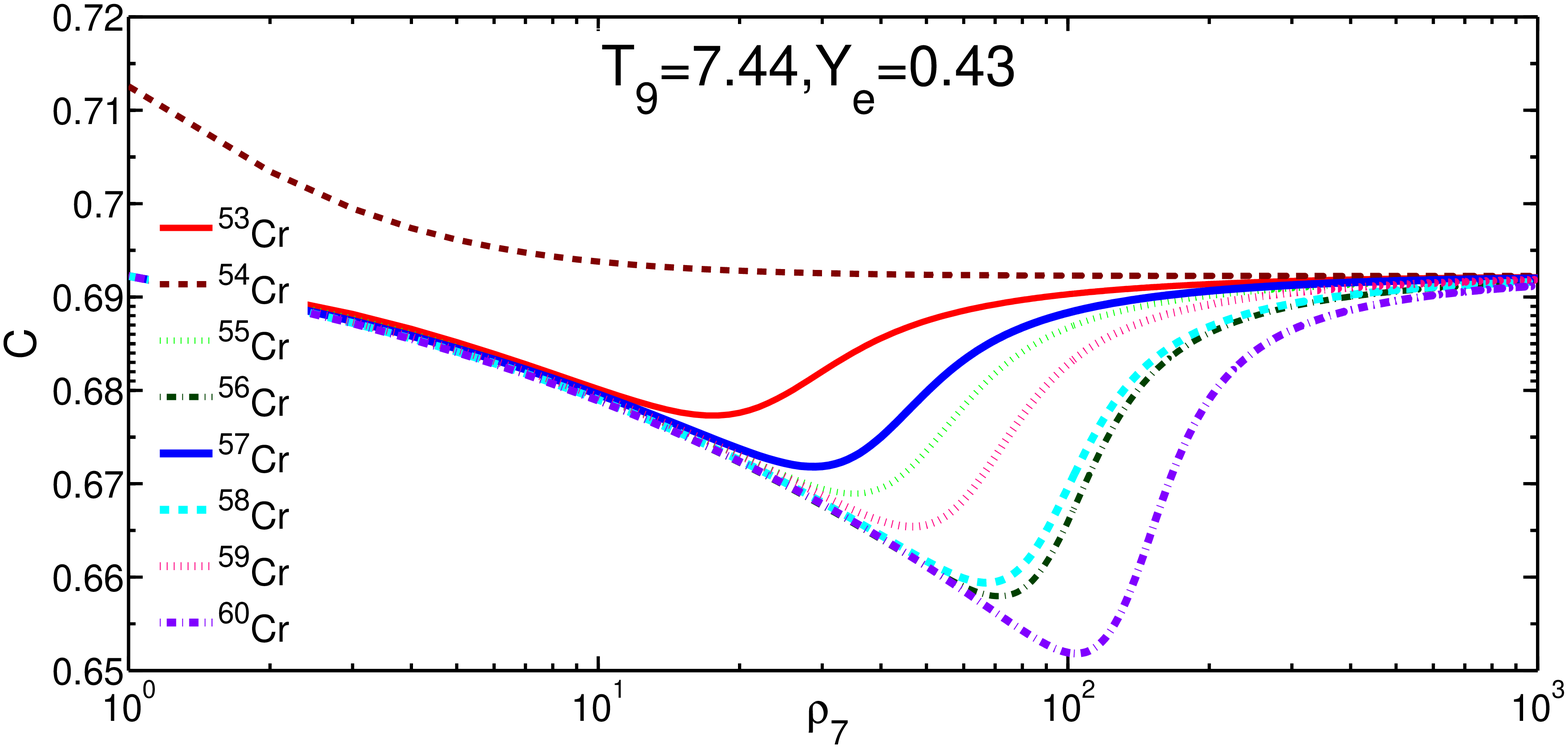}
    \includegraphics[width=7cm,height=7cm]{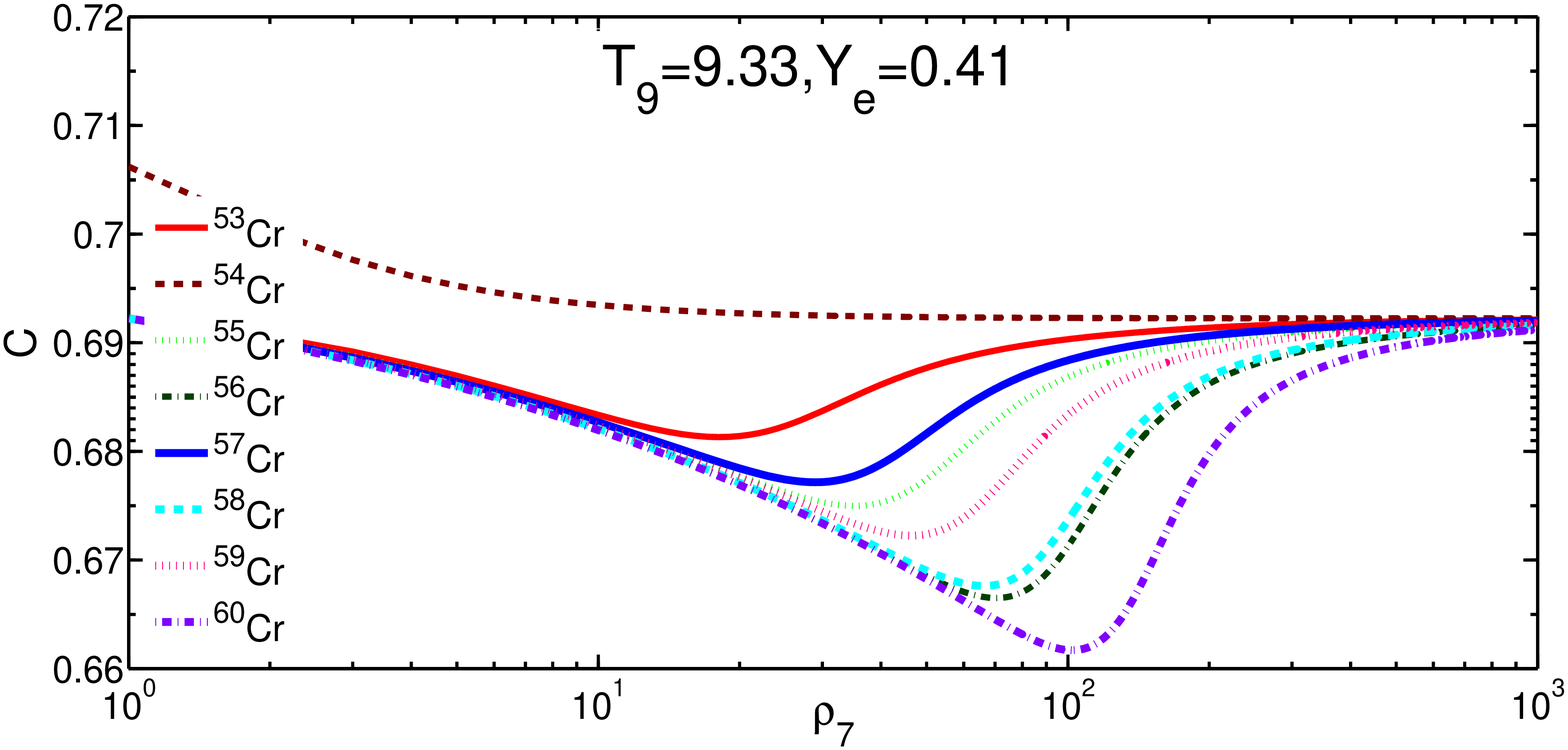}
    \includegraphics[width=7cm,height=7cm]{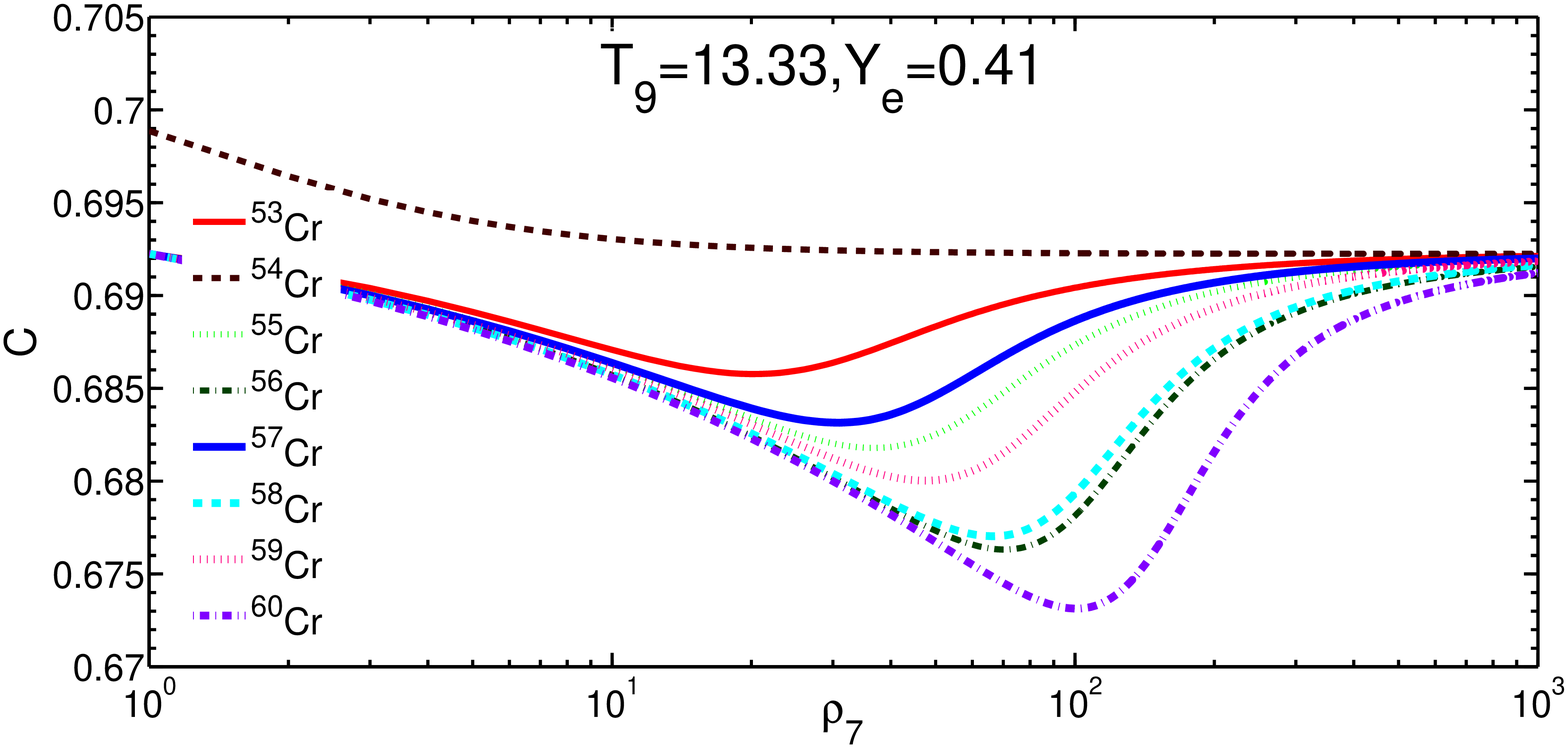}
   \caption{The SES enhancement factor C for chromium isotopes as a function of the density $\rho_7$ at the
temperature of $T_9=3.44,7.44, Y_e=0.43$ and $T_9=9.33, 11.33,
Y_e=0.41$.}
   \label{Fig:2}
\end{figure}

\section{Numerical calculations of EC rates and discussion}
\label{sect:Numerical}

The influences of SES on EC rates for these chromium isotopes at
some typical astrophysics condition are shown in Figure 1. Note that
the no SES and SES rates correspond to solid and dotted line. We
detailed  the EC process according to SMMC method, especially for
the contribution for EC due to the GT transition. For a given
temperature, the EC rates increases by more than six orders of
magnitude as the density increases. Based on proton-neutron
quasiparticle RPA model, Nabi \& Klapdor-Kleingrothaus also detailed
investigated the EC rates in the case without SES. Their results
also shown that the density strongly influence on the EC rates for a
given temperature. For examples, the EC rates for $^{61}$Cr
increases from $6.3096\times10^{-23}\rm{s}^{-1}$ to
$3.71535\times10^2\rm{s}^{-1}$ when the density changes from
$10^7\rm{g/cm}^3$ to $10^{11}\rm{g/cm}^3$ at $T_9=3$ (see the
detailed discussions in Nabi \& Klapdor-Kleingrothaus. 1999). under
the same conditions, the FFN rates for $^{60}$Cr increases from
$8.3946\times10^{-26}\rm{s}^{-1}$ to $1.2388\times10^3\rm{s}^{-1}$
(see Fuller et al. 1982). These studies show that the stellar weak
rates play a key role in the dynamics of the core collapse
calculations and stellar numerical simulation.

According to our calculations, the GT transition EC reaction may not
be dominant process at lower temperature. On the other hand, the
higher the temperature, the larger the electron energy, the larger
the density, the higher the electron Fermi energy becomes.
Therefore, a lot of electrons join in EC reaction and the GT
transition would be very active and have dominated contribution to
total EC rates. Figure 1 shows the screening rates and no screening
rates, which corresponding to solid and dotted line as a function of
density. We find that the screening rates are commonly lower than no
screening rates.

The Gamow-Teller strength distributions play a significant role in
supernova evolution. But the GT$^+$ transitions is addressed only
qualitatively in pre-supernova simulations because of the
insufficient of experimental information. The general rule is that
the energy for the daughter ground state is parameterized
phenomenologically by assuming the GT$^+$ strength resides in a
single resonance. Charge exchange reactions (n, p) and (p, n) would,
if obtainable supply us with plenty of experimental information.
However, any available experimental GT+ strength distributions for
these nuclei can not obtained except for theoretical calculations.
Table 1 present some information about the comparison of our results
by SMMC for total strength, centroid and width of calculated GT
strength distributions with those of NKK (Nabi et al. 2016) for EC
of $^{53-60}$Cr. Our results of GT strength distributions calculated
are higher than those of NKK.

Based the pn-QRPA theory, NKK analyzed nuclear excitation energy
distribution by taking into consideration the particle emission
processes. They calculated stronger Gamow-Teller strength
distribution from these excited states compared to those assumed
using Brink¡¯s hypothesis. However, in their works, they only
discussed the low angular momentum states. By using the method of
SMMC, GT intensity distribution is detailed discussed and actually
an average value of the distribution is adopted in our paper.

The screening factors $C$ is plotted as a function of $\rho_7$ in
figure 2. Due to SES, the rates decrease  by about 40.43\%. The
lower the temperature, the larger the effect of SES on EC rates is.
This is due to the fact that the SES mainly decreased the number of
higher energy electrons, which can actively join in the EC reaction.
Moreover, the SES can also make the EC threshold energy increases
greatly. As a matter of fact, SES will strongly weaken the progress
of EC reactions. One can also find that the screening factor almost
tends to the same value at higher density and it is not dependent on
the temperature and density. The reason is that at higher density
the electron energy is mainly determined by its Fermi energy, which
is strongly decided by density.

Table 2 shows the numerical calculations about the minimum values of
screening factor $C_{\rm{min}}$ in detail. One finds that the EC
rates decrease greatly due to SES. For instance, from Table 2 of the
factor $C_{\rm{min}}$, the rates decrease about 34.75\%, 30.77\%,
36.92\%, 39.07\%, 35.98\%, 38.81\%, 37.50\%, 40.43\% for
$^{53-60}$Cr at $T_9=3.44,Y_e=0.43$, respectively. This is due to
the fact that the SES mainly decreased the number of higher energy
electrons, which can actively join in the EC reactions. On the other
hand, the screening of nuclear electric charges with a high electron
density means a short screening length, which results in a lower
enhancement factor from Coulomb wave correction. However, even a
relatively short electric charge screening length will not have much
effect on the overall rate due to the weak interaction being
effectively a contact potential. A bigger effect is that electrons
are bound in the plasma.

Synthesizes the above analysis, the effects of the charge screening
on the nuclear physics (e.g., EC and beta decay) come at least from
following factors. First, the screening potential will change the
electron Coulomb wave function in nuclear reactions. Second, the
electron screening potential decreases the energy of incident
electrons joining the capture reactions. Third, the electron
screening increases the energy of atomic nuclei (i.e., increases the
single particle energy) in nuclear reactions. Finally, the electron
screening effectively decreases the number of the higher-energy
electrons, whose energy is more than the threshold of the capture
reaction. Therefore, screening relatively increases the threshold
needed for capture reactions and decreases the capture rates.

%

\begin{table*}
\caption{Comparison of our results by SMMC for total strength,
centroid and width of calculated GT strength distributions with
those of NKK (Nabi et al. 2016)for EC of $^{53-60}$Cr.} \centering
 \begin{minipage}{90mm}
  \begin{tabular}{@{}rrrrrrrrrr@{}}
  \hline
 & \multicolumn{2}{c}{$\sum \rm{B(GT)}_+$} & &\multicolumn{2}{c}{$\rm{E}_+$(MeV)}&&\multicolumn{2}{c}{Width$_+$}(MeV)\\
\cline{2-3} \cline{5-6} \cline{8-9} \\
 Nuclide &NKK & SMMC& & NKK & SMMC & & NKK &  SMMC  \\
 \hline
 $^{53}$Cr  &0.51  &0.5625   & &6.21   &6.334   & &2.72   &2.813  \\
 $^{54}$Cr  &1.95  &2.2340   & &2.88   &2.912   & &3.32   &3.406  \\
 $^{55}$Cr  &0.39  &0.4130   & &4.06   &4.126   & &3.47   &3.675  \\
 $^{56}$Cr  &1.31  &1.3326   & &1.77   &1.791   & &2.14   &2.366  \\
 $^{57}$Cr  &0.25  &0.2740   & &5.21   &5.267   & &2.84   &2.972  \\
 $^{58}$Cr  &0.82  &0.8411   & &1.57   &1.605   & &2.49   &2.560   \\
 $^{59}$Cr  &0.24  &0.2520   & &1.26   &1.302   & &2.24   &2.272   \\
 $^{60}$Cr  &0.39  &0.4012   & &3.03   &3.201   & &4.99   &5.017   \\
\hline
\end{tabular}
\end{minipage}
\end{table*}

\begin{table*}
\caption{The minimums value of strong screening factor $C$ for some
typical astronomical condition when $1 \leq\rho_7 \leq 10^3$.}
\centering
 \begin{minipage}{150mm}
  \begin{tabular}{@{}rrrrrrrrrrrr@{}}
  \hline
 & \multicolumn{2}{c}{$T_9=3.44, Y_e=0.43$} & &\multicolumn{2}{c}{$T_9=7.44, Y_e=0.43$}&&\multicolumn{2}{c}{$T_9=9.33,
 Y_e=0.41$}&&\multicolumn{2}{c}{$T_9=13.33, Y_e=0.41$}\\
\cline{2-3} \cline{5-6} \cline{8-9} \cline{11-12}\\
 Nuclide &$\rho_7$ & $C_{\rm{min}}$& & $\rho_7$ & $C_{\rm{min}}$ & &$\rho_7$ & $C_{\rm{min}}$& &$\rho_7$ & $C_{\rm{min}}$  \\
 \hline
 $^{53}$Cr  &18  &0.6525   & &19   &0.6774   & &19   &0.6813     & &20   &0.6858  \\
 $^{54}$Cr  &62  &0.6923   & &65   &0.6924   & &66   &0.6924     & &67   &0.6924      \\
 $^{55}$Cr  &38  &0.6308   & &37   &0.6690   & &36   &0.6750     & &37   &0.6818  \\
 $^{56}$Cr  &81  &0.6093   & &72   &0.6580   & &71   &0.6665     & &71   &0.6763    \\
 $^{57}$Cr  &32  &0.6402   & &30   &0.6719   & &31   &0.6772     & &33   &0.6832  \\
 $^{58}$Cr  &74  &0.6119   & &69   &0.6594   & &67   &0.6676     & &67   &0.6770      \\
 $^{59}$Cr  &50  &0.6250   & &47   &0.6654   & &49   &0.6723     & &48   &0.6800  \\
 $^{60}$Cr  &115 &0.5957   & &106  &0.6518   & &104  &0.6617     & &99   &0.6731    \\
\hline
\end{tabular}
\end{minipage}
\end{table*}

\section{Conclusion remarks}
\label{sect:conclusion}

In this paper, based on the theory of RPA and LRTM, by using the
method of SMMC, we investigated the EC rates in SES. The EC rates
increase greatly by more than six orders of magnitude as the density
increases. On the other hand, by taking into account the influence
of SES on the energy of incident electrons and threshold energy of
electron capture, the EC rates decrease by ~$\sim 40.43$\%.

Electron captures play an important role in the dynamics process of
the collapsing core of a massive star. It is a main parameter for
supernova explosion and stellar collapse. The SES strongly
influences the EC and may influences the cooling rate and
evolutionary timescale of stellar evolution. Thus, the conclusions
we obtained may have a significant influence on the further research
of supernova explosions and numerical simulations.

\begin{acknowledgements}
We would like to thank the anonymous referee for carefully reading
the manuscript and providing some constructive suggestions which are
very helpful to improve this manuscript. This work was supported in
part by the National Natural Science Foundation of China under
grants 11565020, 10773005, and the Counterpart Foundation of Sanya
under grant 2016PT43, the Special Foundation of Science and
Technology Cooperation for Advanced Academy and Regional of Sanya
under grant 2016YD28, the Scientific Research Staring Foundation for
515 Talented Project of Hainan Tropical Ocean University under grant
RHDRC201701, and the Natural Science Foundation of Hainan province
under grant 114012.
\end{acknowledgements}

\label{lastpage}


\begin{thebibliography}{99}

  \bibitem[1990]{aufd90} Aufderheide, M. B., Brown, G. E., kuo, T. T. S., Stout, D. B. and Vogel, P., 1990, ApJ, 362, 241

  \bibitem[1994]{aufd94} Aufderheide, M. B., Fushikii, I., Woosely, S. E. and Hartmanm, D. H., 1994, ApJS, 91, 389

  \bibitem[1999]{Caurier99} Caurier, E.; Langanke, K.; Mart¨ªnez-Pinedo, G.; Nowacki, F., 1999, NuPhA, 653, 439

  \bibitem[1984]{coop84} Cooperstein, J. and Wambach, J., 1984, NuPhA, 420, 591

  \bibitem[1998]{dean98} Dean, D. J., Langanke, K., Chatterjee, L., Radha, P. B., Strayer M. R., 1998, PhRvC, 58, 536

  \bibitem[1980]{full80} Fuller, G. M., Fowler, W. A., Newman, M. J., 1980, ApJS, 42, 447

  \bibitem[1982]{full82} Fuller, G. M., Fowler, W. A., Newman, M. J., 1982, ApJS, 48, 279

  \bibitem[2001]{hege01} Heger, A., Woosley, S. E., Martinez-Pinedo, G. and Langanke, K., 2001, ApJ, 560, 307

  \bibitem[2001]{Itoh02} Itoh, N., Tomizawa, N., Tamamura, M. et al., 2002, ApJ., 579, 380

  \bibitem[1962]{Jancovici62} Jancovici, B., 1962, Nuovo Cimento, 25, 428

  \bibitem[1962]{Jancovici10} Juodagalvis, A., Langanke, K., Hix, W. R., Mart¨ªnez-Pinedo, G., Sampaio, J. M., 2010, NuPhA, 848, 454

  \bibitem[2007]{liu07} Liu, J. J. and Luo, Z. Q., 2007, ChPhy,. 16, 3624

  \bibitem[2008]{liu08} Liu, J. J. and Luo, Z. Q., 2008, CoTP, 49, 239

  \bibitem[2013]{liu13} Liu, J. J., 2013a, MNRAS, 433, 110

  \bibitem[2013]{liu13} Liu, J. J., 2013b, ChPhC, 37, 51018

  \bibitem[2014]{liu14} Liu, J. J., 2014, MNRAS, 438, 930

  \bibitem[2016]{liu16} Liu, J. J., 2016a, RAA, 16, 174

  \bibitem[2016]{liu16} Liu, J. J., and Gu, W. M., 2016b, ApJS, 224, 29

  \bibitem[2017]{liu17} Liu, J. J., 2017, eprint arXiv:1701.05771

  \bibitem[1999]{Nabi99} Nabi, J. and Klapdor-Kleingrothaus, H. V., 1999, eprint arXiv:nucl-th/9907115

  \bibitem[2016]{Nabi16} Nabi, J., Shehzadi, R., and Fayaz, M., 2016, ApSS, 361, 95

  \bibitem[1975]{see75} Seeger, P. A., Howard, W. M., 1975, NuPhA, 238, 491

  \bibitem[2013]{see13} Wanajo, S., Janka, H., M¨¹ller, B., 2013, ApJ, 774, 5


\end{thebibliography}
\end{document}